# 低碳钢填丝脉冲GTAW熔池动态过程多变量Hammerstein模型辨识

吴璟，陈善本

(上海交通大学材料科学与工程学院，上海 200030)

**摘要：** 分析了焊接过程的非线性动态特性，采用多变量 Hammerstein 模型作为辨识对象。采用伪随机序列作为模型输入，用获取的试验数据得到了熔池反面宽度与正面余高的多变量 Hammerstein 模型，并验证了模型的有效性。

**关键词：** GTAW；Hammerstein 模型；非线性系统；DIDO 模型

# DIDO Hammerstein Identification of Mild Steel Welding Pool in Pulsed GTAW Dynamic Process with Wire Filler

Jing Wu, Shan-ben Chen

(School of Materials Science and Eng.，Shanghai Jiaotong Univ., Shanghai 200030, China)

**Abstract:** This paper analyzed the nonlinearity of welding dynamic process, and then adopted MIMO Hammerstein model to describe approximately the process. An identification algorithm was developed and pseudo random signals were adopted as model input. Through a welding experiment, input-output data were obtained and the Hammerstein model of welding pool was identified

  填丝脉冲 GTAW 焊接过程是一个多变量、强耦合、高度非线性的系统，因此有必要建立多变量模型[1]。同时，在焊接过程中熔池可以看作是被焊金属在热输入作用下，经历加热、熔化以及凝固和连续冷却形成的。其中的加热过程本质上是个非线性过程；凝固和冷却过程本质上是个线性过程。[2]所以采用 Hammerstein 模型来描述焊接熔池的动态过程。前人对焊接过程的描述大多集中在线性系统或单变量非线性的模型[3]，难以准确描述复杂的焊接动态过程，本文将建立多变量的 Hammerstein 模型，来描述填丝脉冲 GTAW 焊接过程。

## 1 Hammerstein 模型的结构和参数辨识算法

### 1.1 Hammerstein 模型的表示

  模型的输入变量为：焊接峰值电流 $I_p$ 和送丝速度 $V_f$，输出变量为：背面熔宽 $W_b$ 和正面余高 $H_t$。由于对焊缝成型的控制主要是焊缝熔透，而反映熔透的参数主要是背面熔宽，影响背面熔宽的工艺参数则为焊接线能量 $I_p$ 等参数；同时，本文添加了送丝，填充焊丝影响焊接电源对母材的热输入，引起熔池表面高度的变化。因此选择上述变量作为输入输出量，建立基于 DIDO 的 Hammerstein 模型。

  多变量过程可以用输入输出差分方程

描述成

$$\mathbf{A}(z)\mathbf{y}(k) = \mathbf{B}(z)\mathbf{u}(k) \quad (1)$$

其中矩阵 **A**，**B** 的格式为

$$\begin{cases} \mathbf{C}(z) = \begin{bmatrix} \mathbf{C}_{11}(z) & \mathbf{C}_{12}(z) \cdots & \mathbf{C}_{1m}(z) \\ \mathbf{C}_{21}(z) & \mathbf{C}_{22}(z) \cdots & \mathbf{C}_{21}(z) \\ & \cdots \cdots & \\ \mathbf{C}_{m1}(z) & \mathbf{C}_{m2}(z) \cdots & \mathbf{C}_{mm}(z) \end{bmatrix} \\ \text{其中}\mathbf{C}\text{表示}\mathbf{A}\text{或}\mathbf{B} \end{cases} \quad (2)$$

由于焊接过程是完全可观的，且模型的结构参数可以预先确定，则过程可分解成 m 个子系统。第 s 子系统(s=1,2,…,m)由多项式矩阵 **A**(z)和 **B**(z)对应的第 s 行多项式描述为

$$\begin{cases} \sum_{i=1}^{m} \mathbf{A}_{\mathbf{si}}(z) y_i(k) = \sum_{j=1}^{r} \mathbf{B}_{\mathbf{sj}}(z) u_j(k) \end{cases} \quad (3)$$

式中多项式 $A_{si}$ 和 $B_{sj}$ 如(2)所示。[3]

如果要按照上式进行参数辨识，对应一个子系统需要辨识 m+n+1+p(m+n+2)个参数，显然是很复杂的。而通过观察(1)式可以看出，只要输出变量的系数矩阵 **A**(z)可逆，(1)式即可化为

$$y(k) = \mathbf{A}^{-1}\mathbf{B}x(k) = \mathbf{B}'x(k) \quad (4)$$

其中 $x(k) = \sum_{j=1}^{m} a_j y(k) + \sum_{i=1}^{p} r_i u^i(k) \quad (5)$

将(4),(5)式代入(3)式，(3)式可展成

$$y(k) = -\sum_{i=1}^{n} a_i y(k-i) + \sum_{j=0}^{m} b_j u(k-j) \\ + \sum_{i=2}^{p} \mathbf{B}'(q^{-1})r_i u^i(k) \quad (6)$$

因此，问题的关键在于保证矩阵 A 可逆。我们知道矩阵可逆的充要条件为行列式线形无关，而本文中用于建模的数据为伪白噪声信号，可以满足行列式线形无关的条件。

## 1.2 参数辨识

可将(6)式写成最小二乘格式为

$$\begin{cases} \mathbf{y}_s(k+n_s) = \mathbf{H}_{s0}^{\tau}(k+n_{s0})\boldsymbol{\theta}_{s0} \\ s = 1,2,\cdots,n \end{cases} \quad (7)$$

其中

$$\boldsymbol{\theta}_{s0} = [a_{s1}(1),\cdots r_{1s1}b_{s1}(1),\cdots r_{ps1}b_{s1}^p(1),\cdots]^{\tau}$$

$$\mathbf{H}_{s0} = \begin{bmatrix} \vdots & y_1(k) & \cdots & \vdots & u_1(k) & \cdots & \vdots & u_1^p(k) & \cdots \\ \vdots & y_1(k+1) & \cdots & \vdots & u_1(k+1) & \cdots & \vdots & u_1^p(k+1) & \cdots \\ \vdots & \vdots & & \vdots & \vdots & & \vdots & \vdots & \\ \vdots & y_1(k+L) & \cdots & \vdots & u_1(k+L) & \cdots & \vdots & u_1^p(k+L) & \cdots \end{bmatrix}$$

$$= [\mathbf{y}_1(k),\cdots \mathbf{u}_1(k),\cdots, \mathbf{u}_1(k+n_{s1}-1),\cdots, \mathbf{u}_1^p(k),\cdots, \mathbf{u}_1^p(k+n_{nr}-1)]$$

可用以上推得的最小二乘格式进行递推的参数辨识，得到 $\boldsymbol{\theta}_{s0}$ 值。取

$$\Phi_{N+1} = \begin{pmatrix} \Phi_N \\ \varphi_{N+1} \end{pmatrix}; Y_{N+1} = \begin{pmatrix} Y_N \\ y(k+N+1) \end{pmatrix} \quad (8)$$

记 $P_N = [\Phi_N^T \Phi_N]^{-1}$ 则递推公式为

$$\hat{\boldsymbol{\theta}}_{N+1} = \hat{\boldsymbol{\theta}}_N + \mathbf{P}_N \varphi_{N+1}^T (1 + \varphi_N^T \mathbf{P}_N \varphi_\mathbf{N})^{-1} \\ (y(k+N+1) - \varphi_{N+1}^T \hat{\theta}_N) \quad (9)$$

$$\mathbf{P}_{N+1} = \mathbf{P}_N - \mathbf{P}_N \varphi_{N+1}(1+\varphi_{N+1}^T \mathbf{P}_N \varphi_{\mathbf{N+1}})^{-1} \varphi_{N+1}^T \mathbf{P}_N \quad (10)$$

## 1.3 模型的结构辨识[4]

对(7)式的结构辨识，是指对线性子系统的阶次 n, m 和非线性环节的最高幂次 p 的辨识，对于延迟时间，通过比较输入输出数据得到延迟时间 $\tau_{d1}=1, \tau_{d2}=3$。

考虑到多变量辨识时的数据量非常大，因此本文采用递推的方法进行结构辨识。

首先对非线性环节的阶次 p 进行辨识：如取 p=p₀+1 时，相对应的数据矩阵增加到 n+1 列，对应的辨识方程为

$$y_s = \begin{bmatrix} \mathbf{H}_1 & \mathbf{H}_2 \end{bmatrix} \begin{bmatrix} \boldsymbol{\theta}_{01} \\ \boldsymbol{\theta}_{02} \end{bmatrix} \quad （11）$$

式中 **H₁**, **θ₀₁** 与(7)式定义的 **H₀**, **θ₀** 的维数相同，而且 **H₁**= **H₀**,

$$\mathbf{H}_2 \triangleq [\mathbf{u}_1^{p_0+1}(k) \cdots \mathbf{u}_1^{p_0+1}(k+n_1)\cdots \mathbf{u}_r^{p_0+1}(k) \cdots \mathbf{u}_r^{p_0+1}(k+n_r)]$$

$$\boldsymbol{\theta}_{02} = [r_{p_0+1,s}b_{1,s} \cdots r_{p_0+1,s}b_{1,s}\cdots r_{p_0+1,s}b_{r,s} \cdots r_{p_0+1,s}b_{r,s}] \quad (12)$$

式(7)的最小二乘解为

$$\begin{aligned}\hat{\boldsymbol{\theta}}_{01} &= \hat{\boldsymbol{\theta}}_0 - \mathbf{A}\boldsymbol{\Phi}_2^T(\mathbf{y} - \boldsymbol{\Phi}_1\hat{\boldsymbol{\theta}}_0) \\ \hat{\boldsymbol{\theta}}_{02} &= \mathbf{B}\boldsymbol{\Phi}_2^T(\mathbf{y} - \boldsymbol{\Phi}_1\hat{\boldsymbol{\theta}}_0)\end{aligned} \quad (13)$$

式中 $\mathbf{A} = (\boldsymbol{\Phi}_1^T\boldsymbol{\Phi}_1)^{-1}\boldsymbol{\Phi}_1^T\boldsymbol{\Phi}_2\mathbf{B}$
$\mathbf{B} = (\boldsymbol{\Phi}_2^T\boldsymbol{\Phi}_2 - \boldsymbol{\Phi}_2^T\boldsymbol{\Phi}_1(\boldsymbol{\Phi}_1^T\boldsymbol{\Phi}_1)^{-1}\boldsymbol{\Phi}_1^T\boldsymbol{\Phi}_1)^{-1}$

在初始估计 $\hat{\boldsymbol{\theta}}_0$ 之下，由式(12)构成了对 $p=p_0, p_0+1\ldots$ 的递推辨识算法，于是每次运算只需对 $\sum_{i=0}^{r}(n_{si}+1) \times \sum_{i=0}^{r}(n_{si}+1)$ 维矩阵 $\mathbf{B}$ 求逆，这比直接求 $(\boldsymbol{\Phi}_1^T\boldsymbol{\Phi}_1)^{-1}$ 运算量小得多。在获得一个 $\hat{\theta}(p)$ 后，代入指标函数 $J[\hat{\theta}(p)]$ 中，计算不同的 p 值时的 J 值，选取 J 随 p 值增加不明显减小时的 $\hat{p}$ 值作为估计值。

线性子系统阶次 n, m 的估计与 p 的估计方法类似，此处不予赘述。

递推初值可取为：$\begin{cases}\hat{\theta}_0 = 0 \\ p_0 = \alpha^2 I, \quad 10^5 \leq \alpha^2 \leq 10^{10}\end{cases}$

## 2 焊接过程模型辨识

本文选择了背面熔宽 $W_b$ 和正面余高 $H_t$ 作为建模输出量，焊接峰值电流 $I_p$ 和送丝速度 $V_f$ 为输入量，建立描述填丝脉冲 GTAW 过程的 DIDIO 模型。共采集了 1070 个数据，其中 1000 个数据用于建模，70 个数据用于模型的检验。

### 2.1 辨识激励信号的选择

为了对非线性动态模型进行辨识，输入信号必须充分激励过程的所有模态，并且为了消除 $\Phi^T\Phi$ 的奇异性,理想的输入信号应是高斯白噪声，但这种信号的产生在技术上由一定的困难，因此我们用伪随机噪声作为试验输入信号，产生的方法采用递推同余法。对图 1 所示的模型进行辨识，得到了较理想的结果。

因此本文分别以峰值电流和送丝速度的随机白噪声输入信号作为激励信号，其变化范围和变化步长分别确定为：$I_p$=(130~170A,2A)，$V_f$=(4~10cm/s,1cm/s)，基本工作点为：$I_{p0}$=150A，$V_{w0}$ =1.9mm/s，$V_{f0}$=7.0cm/s，工件接头形式为对接，采用脉冲送丝方式，采样时间为 1$s$。[5]

受系统工作方式和干扰因素的影响，输入输出数据中通常含有直流成分和高低频噪声，这些将会严重的影响辨识结果，因此要对实验的输入输出数据进行了中值滤波处理和去除直流成分。

### 2.2 焊接过程模型辨识结果

采用 RLS 对模型结构辨识采用误差 函数检验法，取目标函数 $J(\hat{\theta})$ 随着模型结构参数 n，m, p 的增加时减小趋于平稳的值。从辨识实验结果可以看出，目标函数 $J(\hat{\theta})$ 的值分别在 p=2,m=5,n=3（第一个输出）和 p=4,m=5,n=5（第二个输出）趋于稳定下降。由于考虑到用于控制的模型不易于过于复杂，最终确定基于 Hammerstein 的 DIDO 模型结构参数分别为：第一个输出：$d$=1, $p$=2, $m$=5, $n$=3；第二个输出：$d$=3, $p$=4, $m$=5, $n$=5。辨识结果为，模型结构见图 1：

$$f_{11}(u) = u - 0.01476u^2, \qquad f_{21}(u) = u - 0.04142u^2$$

$$f_{12}(u) = u + 0.002972u^2 - 0.00315u^3 + 0.000152u^4$$

$$f_{22}(u) = u + 0.115034u^2 + 0.133773u^3 - 0.02614u^4$$

$$\frac{\mathbf{B}_{11}}{\mathbf{A}_{11}} = \frac{q^{-1}(0.004744 - 0.0031q^{-1} + 0.000158q^{-2} - 0.0015q^{-3})}{1 - 1.73603q^{-1} + 0.728305q^{-2} + 0.580712q^{-3} - 0.85552q^{-4} + 0.320009q^{-5}}$$

$$\frac{\mathbf{B}_{21}}{\mathbf{A}_{11}} = \frac{q^{-3}(0.001614 - 0.0047q^{-1} - 0.00742q^{-2} + 0.0000138q^{-3} - 0.00924q^{-4} + 0.002941q^{-5})}{1 - 1.73603q^{-1} + 0.728305q^{-2} + 0.580712q^{-3} - 0.85552q^{-4} + 0.320009q^{-5}}$$

$$\frac{\mathbf{B}_{12}}{\mathbf{A}_{22}} = \frac{q^{-1}(0.00568 + 0.002351q^{-1} + 0.000844q^{-2} + 0.000724q^{-3} - 0.00253q^{-4} - 0.00333q^{-5})}{1 - 1.29125q^{-1} + 0.253601q^{-2} + 0.543266q^{-3} - 0.69655q^{-4} + 0.240607q^{-5}}$$

$$\frac{\mathbf{B}_{22}}{\mathbf{A}_{22}} = \frac{q^{-3}(0.005929 - 0.01733q^{-1} + 0.010646q^{-2} - 0.01391q^{-3} - 0.00406q^{-4} - 0.02969q^{-5})}{1 - 1.29125q^{-1} + 0.253601q^{-2} + 0.543266q^{-3} - 0.69655q^{-4} + 0.240607q^{-5}}$$

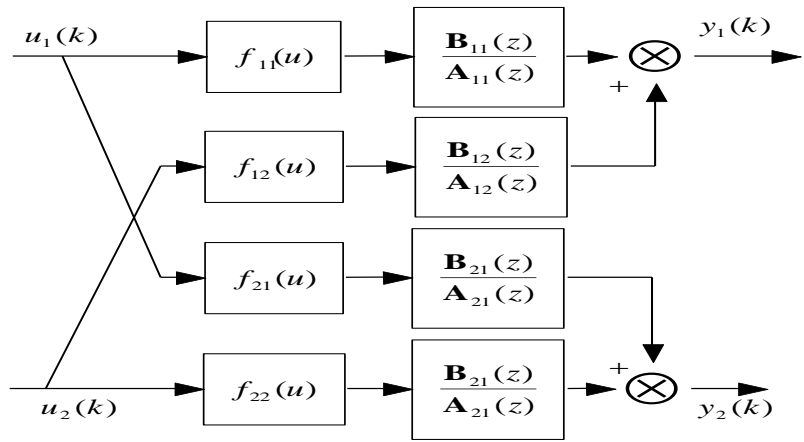

图 1 Hammerstein 模型

Fig.1 Hammerstein mod

### 2.3 模型的检验

为了考察模型的性能，利用剩余的 70 个实际观测数据对上述模型进行测试。方法是将测试样本实际值与模型输出值比较，观察误差变化。误差统计结果为：第一个输出的平均误差和标准方差分别为 0.07973 和 0.07769；第二个输出的平均误差和标准方差分别为-0.07977 和 0.03096。可以看出模型的误差完全符合焊接控制的要求[6-10]，从而进一步证明了基于 Hammerstein 模型的焊接过程建模的精确性。

## 3 结论

本文基于 GTAW 熔池动态过程建立了多变量 Hammerstein 模型，从模型的检验来看，系统误差小，说明了用多变量 Hammerstein 模型描述动态的焊接过程有效和可行的。说明用多变量 Hammerstein 模型描述焊接过程比用一般的线性模型更准确。这也为下一步基于模型结合模糊和神经网络等技术设计控制器奠定了基础。

**参考文献**

[1] S. Chen, J. Wu, "Intelligentized Methodology for Arc Welding Dynamical Processes", Springer, Berlin Heidelberg, 2009

[2] 张文钺，焊接传热学，北京，机械工业出版社,1987:18-33.

[3] 方崇智，萧德云，过程辨识，北京，清华大学出版社，1988，84-92.

[4] 高玉琦，陈善本，用Hammerstein模型描述的非线性系统的结构辨识[J] 机器人，1994, 16(1)1-6

[5] S. B. Chen, J. Wu, Q. Y. Du, "Non-linear modelling and compound intelligent control of pulsed gas tungsten arc welding dynamics." Proc IME J Syst Contr Eng, Part I: Journal of Systems and Control Engineering, vol. 225, no. 1, pp. 113-124, 2011.

[6] S. Chen, J. Wu, "Intelligentized Technology for Arc Welding Dynamical Processes", Springer, Berlin (LNEE 29), 2009

[7] W. Li, J. Wu, et al, "Rough set based modeling for welding groove bottom state in narrow gap MAG welding", Industrial Robot: An International Journal, Vol. 42 Iss: 2, pp.110 – 116, 2015

[8] H. Shen, J. Wu, T. Lin, S. Chen, "Arc Welding Robot System with Seam Tracking and Weld Pool Control Based on Passive Vision", Int. J. Adv. Manuf. Technol., vol. 39, no. 7-8, pp. 669-678, 2007

[9] S. Chen, J. Wu, "A Survey on Intelligentized Technologies for Visual Information Acquirement, modeling and Control of Arc Welding Pool Dynamics", Proc. 33rd Annual Conference of the IEEE Industrial Electronics Society (IECON'07), Taipei, Taiwan, R.O.C., 2007.

[10] W. Li, J. Wu, et al, "Rough set based modeling for welding groove bottom state in narrow gap MAG welding", Industrial Robot: An International Journal, Vol. 42 Iss: 2, pp.110 – 116, 2015